%% file: main.tex
\documentclass[conference]{IEEEtran}

\input confpaperdefs.tex

\begin{document}



\title{\LARGE \textbf{Removing Time-Scale Separation in Feedback-Based Optimization via Estimators}}




\author{
    \IEEEauthorblockN{Niloufar Yousefi and John W. Simpson-Porco}
}

\maketitle

\begin{abstract}
 Feedback-based optimization (FBO) provides a simple control framework for regulating a stable dynamical system to the solution of a constrained optimization problem in the presence of exogenous disturbances, and does so without full knowledge of the plant dynamics. However, closed-loop stability requires the controller to operate on a sufficiently slower timescale than the plant, significantly constraining achievable closed-loop performance. Motivated by this trade-off, we propose an estimator-based modification of FBO which leverages dynamic plant model information to eliminate the time-scale separation requirement of traditional FBO. Under this design, the convergence rate of the closed-loop system is limited only by the dominant eigenvalue of the open-loop system. We extend the approach to the case of design based on only an approximate plant model when the original system is singularly perturbed. The results are illustrated via an application to fast power system frequency control using inverter-based resources.
\end{abstract}

\section{Introduction}\label{Sec:Introduction}

The growing complexity of modern engineered systems \textemdash{} such as energy systems \cite{IF-EE-JWSP-EF-MP-AH:24a, ZW-WW-JZFP-FL-BY:23}, communication networks \cite{YT-ZW-XY-XS-JY-HZ:25}, robotic teams \cite{AO-MT-AS-DL-AF:22}, and supply chains \cite{TYC-DRK:06} \textemdash{}  has created a need for flexible control paradigms capable of rejecting disturbances, enforcing operating constraints, and optimizing system performance. Common techniques achieving these objectives include model (or data-driven) predictive control \cite{JBR-DQM:09,JBR-DA-CNB:12,JC-JL-FD:19}, extremum seeking \cite{AS:24}, and our focus here, \emph{feedback-based optimization} (FBO) \cite{AH-ZH-SB-GH-FD:24}, also known as online feedback optimization \cite{MP-LO-SB-FD:22} or optimal steady-state control \cite{LSPL-JWSP-EM:18l}. 

FBO operates by embedding simple optimization elements directly into the controller for the dynamical system \cite{MAH-EMU-MEB-24, DKM-FD-HS-SHL-SC-RB-JL:17, GB-JC-JIP-ED:21, MC-EDA-AB:18, AH-SB-GH-FD:21, AH-ZH-SB-GH-FD:24, LSPL-ZEN-EM-JWSP:18e,ED-AS:16, AB-ED-AS:18}, in order to continuously steer the system toward an optimal equilibrium \cite{MC-EDA-AB:18}. FBO leverages real-time system measurements, and uses only limited model information, in particular the steady-state input-to-output mapping of the plant. This can be viewed as a generalization of classical integral control, shifting from reference tracking to direct optimization of economic or operational objectives \cite{AH-ZH-SB-GH-FD:24}.

Despite these advances, a persistent challenge in FBO is ensuring closed-loop stability when the physical plant and the optimization dynamics operate on similar time scales. Typical stability results (e.g., \cite{SM-AH-SB-GH-FD:18, AH-SB-GH-FD:21,MC-EDA-AB:18,AH-ZH-SB-GH-FD:24, GB-JC-JIP-ED:21}) require a separation of time scales between the plant and the controller, which in practice is enforced by reducing the controller gain. These results will be reviewed in Section~\ref{Sec:ReviewFBO}. Unfortunately, this low-gain requirement leads to performance limitations, and restricts the use of FBO to applications where either (a) the dynamics are fast, or (b) slow control is acceptable; such applications power systems \cite{MC-EDA-AB:18,ZT-EE-JWSP-EF-MP-HH:20l,MP-LO-SB-FD:22} and building automation \cite{GB-DLM-MHDB-SB-RS-JL-FD:24}.

The ability to operate an FBO controller on the same timescale as the plant would improve transient performance and broaden the range of practical FBO applications. This motivates the study of (i) conditions under which FBO stability can be guaranteed without timescale separation, and (ii) the design of entirely new FBO methods. Pursuing the first direction, the work in \cite{MB-FD:25} removes the timescale separation requirement by imposing, roughly speaking, a block-diagonal dominance condition on the closed-loop Jacobian; this condition guarantees global exponential stability for any FBO gain. In a complementary result, the same authors in \cite{MB-FD} analyze monotone dynamical plants and establish global convergence using a small-gain argument on steady-state input–output maps. These conditions depend only on steady-state sensitivities rather than full dynamic models. However, the assumptions in \cite{MB-FD:25,MB-FD} may restrict their general applicability. Regarding the second direction, to our knowledge the only existing work is \cite{LSPL-JWSP-EM:18l}, which proposes a design that does not require timescale separation but is limited to quadratic objectives and lies somewhat outside the current FBO literature. 

\emph{Contributions:} We develop here a new and generally applicable FBO design for LTI systems to remove the limitation of timescale separation. Our approach introduces an estimator into the loop, based on a dynamic model of the plant. The estimator generates a real-time prediction of the plant’s idealized steady-state output, given the current instantaneous value of the input. Remarkably, injecting this estimate signal into a standard FBO design eliminates the timescale separation requirement. 
Section~\ref{Sec:EstFBO} describes our design procedure for the case where full plant model information is available. To reduce the procedure's dependence on exact plant models, in Section~\ref{Sec:EstFBO} we extend to the case of two-timescale system models, wherein the proposed estimator is designed based only on the slow timescale subsystem model. In both cases, we provide rigorous theoretical results establishing stability without timescale separation between the plant and controller. The results are illustrated in Section \ref{Sec:Example} through an application to power system frequency control with inverter-based resources, where the EE-FBO design significantly enhances closed-loop performance. Our approach opens up a new direction in FBO design, where incorporated model information can be traded off against closed-loop performance in a systematic fashion. 




\section{Background: Feedback-Based Optimization via Gradient Flow}\label{Sec:ReviewFBO}
Consider the LTI system given by 
\begin{equation}\label{Eq:LTI}
    \begin{aligned}
        \dot{x} &= Ax +Bu +Ew\\
        y &= Cx
    \end{aligned}
\end{equation}
with state $x(t) \in \real^{n}$, control input $u(t) \in \real^m$, constant disturbance $w \in \mathbb{R}^q$ and measured output $y(t) \in \mathbb{R}^p$. As is standard in FBO \cite{AH-ZH-SB-GH-FD:24}, we assume that the open-loop plant is asymptotically stable, i.e., $A$ is Hurwitz. As a consequence, for any fixed input $u(t) = \bar{u}$, the state of \eqref{Eq:LTI} converges towards the unique equilibrium value $\bar{x} = -A^{-1}(B\bar{u}+Ew)$, and the equilibrium output is given by
\begin{equation}\label{Eq:SSIO}
    \bar{y}=\Pi_u \bar{u}+\Pi_w w,
\end{equation}
where $\Pi_u \define -CA^{-1}B \in \mathbb{R}^{p \times m}$ and $\Pi_w \define -CA^{-1}E \in \mathbb{R}^{p \times q}$. The problem of interest is to design a controller which drives the output and control input of \eqref{Eq:LTI} towards the solution of the equilibrium-constrained optimization problem
\begin{equation}\label{Eq:optimization}
    \begin{aligned} 
        \minimize_{\bar{u} \in \mathcal{U}} \quad & f(\bar{u})+g(\bar{y})\\
        \subto \quad & \bar{y}=\Pi_u \bar{u} +\Pi_w w
    \end{aligned}
\end{equation}
where $\mathcal{U} \subset \real^m$ is an input constraint set and $f : \mathbb{R}^m \rightarrow \mathbb{R}$ and $g : \mathbb{R}^p \rightarrow \mathbb{R}$ are costs placed on the equilibrium input and output values, respectively. We place the following technical assumptions on the problem data.

\smallskip

\begin{assumption}[\bf Cost Assumptions]\label{Ass:Cost}
    The set $\mathcal{U}$ is closed, non-empty, and convex. The input cost function $f$ is continuously differentiable, $\mu_f$-strongly convex, and its gradient $\nabla f$ is globally $\ell_f$-Lipschitz continuous, where $\mu_f, \ell_f > 0$. The output cost function $g$ is continuously differentiable, convex, and its gradient $\nabla g$ is globally $\ell_{g}$-Lipschitz continuous. \hfill \oprocend
\end{assumption}

\smallskip

Assumption \ref{Ass:Cost} can be significantly relaxed (e.g., \cite{SM-AH-SB-GH-FD:18}) but is appropriate for our purposes here. Under Assumption \ref{Ass:Cost}, for each $w \in \real^q$ the problem \eqref{Eq:optimization} possesses a unique minimizer $(\bar{u}^{\star},\bar{y}^{\star})$. Note that the linear equality constraint in \eqref{Eq:optimization} can be eliminated, yielding the unconstrained problem
\begin{equation}\label{optimization_simple}
        \min_{\bar{u} \in \mathcal{U}} \quad f(\bar{u})+g(\Pi_u \bar{u} +\Pi_w w),
\end{equation} 
which may then be solved from any initial condition $u(0)$ via the \emph{projected gradient flow}
\begin{equation}\label{Eq:ProjectedGradientFlow}
    \tau \dot{u} = P_{\mathcal{U}}\Big(u- \eta \big( \nabla f(u) +  \Pi_u^{\T} \nabla g(\Pi_u u +\Pi_w w)\big)\Big)-u
\end{equation}
where $\eta,\tau  > 0$ are tuning parameters, and $\mathcal{P}_{\mathcal{U}}$ denotes the Euclidean projection onto $\mathcal{U}$ \cite{SB-LV:04}. We adopt the continuous (Lipschitz) projection formulation~\cite{GB-JC-JIP-ED:21}, which preserves feasibility while avoiding the discontinuities of classical projected dynamical systems~\cite{YSX-JW:00}. For appropriately chosen $\eta$, the algorithm~\eqref{Eq:ProjectedGradientFlow} is unconditionally stable, and its convergence rate can be increased arbitrarily by decreasing $\tau$.

A feedback controller can be obtained from \eqref{Eq:ProjectedGradientFlow} by replacing the steady-state output value $\bar{y} = \Pi_u u +\Pi_w w$ with the \emph{measured} output value of $y(t)$, yielding the controller
\begin{equation}\label{Eq:GradientController}
\tau \dot{u} = P_{\mathcal{U}}\Big(u- \eta \Phi(u,y) \Big)-u,
\end{equation}
where
\begin{equation}\label{Eq:Phi}
\Phi(u,y) \define  \nabla f(u) +  \Pi_u^{\T} \nabla g(y).
\end{equation}
An advantage of the controller \eqref{Eq:GradientController}--\eqref{Eq:Phi} is that it requires only the DC gain $\Pi_u$ of the plant \eqref{Eq:LTI} as model information. The following result can be proved via small-gain or singular perturbation methods (e.g., \cite{SM-AH-SB-GH-FD:18, JWSP:20a}). 

\smallskip

\begin{theorem}[\bf Low-Gain Stability of FBO]\label{Thm:FBO}
Assume that $A$ is Hurwitz, that Assumption \ref{Ass:Cost} holds, and select $\eta \in (0,2\mu_{f}/(\ell_f + \ell_g \|\Pi_u\|_{2}^2)^2)$. Then (i) for each $w \in \real^{q}$, the closed-loop system \eqref{Eq:LTI}, \eqref{Eq:GradientController} possesses a unique equilibrium point $(\bar{x},\bar{u}^{\star})$ with corresponding output $\bar{y}^{\star}$, and (ii) there exists $\tau^{\star} > 0$ such that the equilibrium point is globally exponentially stable for all $\tau > \tau^{\star}$.
\end{theorem}

\smallskip

 In contrast with the underlying pure gradient flow \eqref{Eq:ProjectedGradientFlow} on which the controller \eqref{Eq:GradientController} is based, stability of the closed-loop system \eqref{Eq:LTI}, \eqref{Eq:GradientController} is not unconditional, and the minimum theoretical value $\tau^{\star}$ of $\tau$ guaranteeing stability can be very large, leading to poor closed-loop performance.


\section{Estimator-Based Performance Enhancement of FBO}
\label{Sec:EstFBO}

 In light of the unconditional stability of the gradient flow \eqref{Eq:ProjectedGradientFlow}, one way to understand the poor achievable performance of the controller \eqref{Eq:ProjectedGradientFlow} is simply that the measured output $y(t)$ used in the controller \eqref{Eq:ProjectedGradientFlow} differs from the steady-state value $\bar{y} = \Pi_u u + \Pi_w w$ during transients. It is this steady-state output value the gradient flow ``expects'' to receive, and this discrepancy leads to the performance limitation. To remedy this problem, further model information can be used to produce an \emph{estimate} $\hat{\bar{y}}$ of the steady-state output, which can be fed into the controller \eqref{Eq:ProjectedGradientFlow}. We next describe how this can be achieved via an estimator \emph{if} the dynamic model \eqref{Eq:LTI} is available.


\subsection{Estimator Design Based on Full Plant Model}
\label{Sec:EstFBOMotivation}

Since the disturbance $w$ is constant, it is described by the simple differential equation $\dot{w} = 0$. Appending this to the model \eqref{Eq:LTI}, we obtain the augmented system
\begin{equation}\label{Eq:Augmented}
\begin{aligned}
\begin{bmatrix}
\dot{x}\\
\dot{w}
\end{bmatrix} &= \underbrace{\begin{bmatrix}A & E \\ 
\vzeros & \vzeros
\end{bmatrix}}_{\define \mathbf{A}_{\rm aug}}\begin{bmatrix}x \\ w\end{bmatrix} + \underbrace{\begin{bmatrix}B \\ \vzeros \end{bmatrix}}_{\define \mathbf{B}_{\rm aug}}u, \quad y = \underbrace{\begin{bmatrix}C & \vzeros \end{bmatrix}}_{\define \mathbf{C}_{\rm aug}}\begin{bmatrix}x \\ w\end{bmatrix}.
\end{aligned}
\end{equation}
To design an estimator for the augmented model \eqref{Eq:Augmented}, we require detectability of $(\mathbf{A}_{\rm aug},\mathbf{B}_{\rm aug})$, which since $A$ is Hurwitz is ensured by the following assumption.

\smallskip

\begin{assumption}[\bf Non-Resonance]\label{Ass:NonRes}
The matrix $\left[\begin{smallmatrix}
    A & E \\
    C & \vzeros
\end{smallmatrix}\right]$ has rank $n+q$, i.e., full column rank.
\end{assumption}

\smallskip

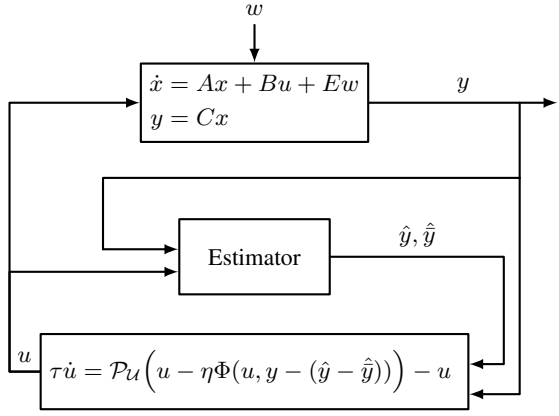
\begin{figure}[ht!]
    \centering
    \input{Figures/fig1}
    \caption{Feedback-based optimization with an estimator. The estimator predicts output and its steady state, feeding these into the controller.}
    \label{Fig:Estimator}
\end{figure}

Our proposed estimator-enhanced feedback-based optimization (EE-FBO) design, replacing  \eqref{Eq:GradientController}, is 

\begin{subequations}\label{Eq:FBOEstimator}
    \begin{align}
    \label{Eq:FBOEstimator-1}
    \begin{bmatrix}
    \dot{\hat{x}}\\
    \dot{\hat{w}}
    \end{bmatrix} &= \mathbf{A}_{\rm aug}\begin{bmatrix}\hat{x} \\ \hat{w}\end{bmatrix} + \mathbf{B}_{\rm aug}u - \mathbf{L}(y - \hat{y})\\
    \label{Eq:FBOEstimator-3}
    \hat{y} &= \mathbf{C}_{\rm aug}\begin{bmatrix}\hat{x} \\ \hat{w}\end{bmatrix}\\
    \label{Eq:FBOEstimator-4}
    \hat{\bar{y}}&=\Pi_u u+\Pi_w \hat{w}\\
    \label{Eq:FBOEstimator-5}
    \tau \dot{u} &= P_{\mathcal{U}}\Big(u- \eta \Phi(u,y-(\hat{y}-\hat{\bar y}))\Big)-u,
    \end{align}
\end{subequations}
where $\mathbf{L} \in \real^{(n + q) \times p}$ is an estimator gain to be designed,  $\eta, \tau > 0$, and $\Phi$ is as defined in \eqref{Eq:Phi}. Figure \ref{Fig:Estimator} shows a block diagram of the design.

The estimator generates two predictions: (i) the current output $\hat{y}$ in \eqref{Eq:FBOEstimator-3}, and (ii) the steady-state value $\hat{\bar{y}}$ in \eqref{Eq:FBOEstimator-4} that would occur \emph{if} $u$ and $\hat{w}$ were constant. The estimated deviation from the steady-state is calculated as $\hat{y} - \hat{\bar{y}}$, which is subtracted from the true measurement $y$ in the gradient control law \eqref{Eq:FBOEstimator-5}. 


\subsection{Stability Analysis}

We now study the stability of the closed-loop system \eqref{Eq:LTI}, \eqref{Eq:FBOEstimator}. The following technical result will be useful in the subsequent analysis and in Section \ref{Sec:ReducedEstFBO}; see Appendix \ref{Sec:Proof} for the proof.

\smallskip

\begin{proposition}[\bf Exponential ISS of a Projection Algorithm]\label{Prop:ControllerISS} 
Let $\mathcal{U} \subset \real^m$ be a closed non-empty convex set, let $\map{\pi}{\real^m \times \real^p \times \real^{n_2}}{\real^m}$, and consider the dynamics
\begin{equation}\label{Eq:CTProjection}
\dot{u} = P_\mathcal{U}(u - \eta \pi(u, v,e)) - u,
\end{equation}
where $\eta > 0$ and $(v,e)$ are external input signals. Suppose that $\pi$ is (i) \( \mu \)-strongly monotone and \( L \)-Lipschitz continuous in \( u \), uniformly in $(v,e)$, and (ii) $L_v'$-Lipschitz in $v$ and $L_e'$-Lipschitz in $e$, uniformly in the other arguments. Denote by \( u^* \in \mathcal{U} \) the unique solution to \( u^* = P_\mathcal{U}(u^* - \pi(u^*, 0, 0)) \). 

Then for $\eta < 2\mu/L^2$, \eqref{Eq:CTProjection} is exponentially input-to-state stable (ISS) with respect to the equilibrium $u^*$, and the ISS gains $\gamma_{v\to u}$ and $\gamma_{e\to u}$ with respect to $v$ and $e$ can be taken as 
\begin{equation}\label{Eq:ControllerL2Gains}
\gamma_{v \to u} = \frac{\eta L_v'}{1-\rho}, \quad  \gamma_{e \to u} = \frac{\eta L_e'}{1-\rho},
\end{equation}
where \( \rho = \sqrt{1 - 2\eta \mu + \eta^2 L^2} \in (0, 1) \).
\end{proposition}

\smallskip

We can now state the first main result, showing that the design \eqref{Eq:FBOEstimator} removes the tuning limitations of \eqref{Eq:GradientController}.

\smallskip

\begin{theorem}[\bf Unconditional Stability of EE-FBO]\label{Thm:EEFBO}
Assume that $A$ is Hurwitz and that Assumptions \ref{Ass:Cost} and \ref{Ass:NonRes} hold. Select $\eta \in (0,2\mu_{f}/(\ell_f + \ell_g \|\Pi_u\|_{2}^2)^2)$, and select $\mathbf{L}$ such that $\mathbf{A}_{\rm aug} + \mathbf{L}\mathbf{C}_{\rm aug}$ is Hurwitz. Then (i) for each $w \in \real^{q}$, the closed-loop system \eqref{Eq:LTI}, \eqref{Eq:FBOEstimator} possesses a unique equilibrium point $(x,\hat{x},\hat{w},u) = (\bar{x},\bar{x},w,\bar{u}^{\star})$ with corresponding output $\bar{y}^{\star}$, and (ii) the equilibrium point is globally exponentially stable \underline{for all} $\tau > 0$. 
\end{theorem}

\smallskip

\begin{pfof}{Theorem \ref{Thm:EEFBO}}
Applying the PBH test \cite[Theorem 14.2]{JPH:09} and using Assumption \ref{Ass:NonRes}, it is straightforward to show that \eqref{Eq:Augmented} is detectable, and thus $\mathbf{L}$ can be selected such that $\mathcal{A}\define \mathbf{A}_{\rm aug} + \mathbf{L}\mathbf{C}_{\rm aug}$ is Hurwitz. Define the error variables $\tilde{\zeta}\define \begin{bmatrix}
    x-\hat{x} & w-\hat{w}
\end{bmatrix}^{\sf T}$, and $\Tilde{y} \define y-\hat{y}$. In the new coordinates, the estimator dynamics \eqref{Eq:FBOEstimator-1}--\eqref{Eq:FBOEstimator-4} become
\begin{equation}\label{Eq:EstErrorDynamics}
\begin{aligned}
\dot{\tilde{\zeta}} &= \mathcal{A}\tilde{\zeta}\\
\Tilde{y}&=\mathbf{C}_{\rm aug}\tilde{\zeta}\\
        \hat{\bar{y}} &= \Pi_u u+\Pi_w(w-\Tilde{w}).
\end{aligned}
\end{equation}
and the controller \eqref{Eq:FBOEstimator-5} similarly becomes
\begin{equation}\label{Eq:EstErrorController}
    \begin{aligned}
        \tau \dot{u}&=P_{\mathcal{U}}\big(u- \eta (\nabla f(u) +\Pi_u^{\sf T} \nabla g(\tilde{y}+\hat{\bar{y}}))\big)-u\\
        &=P_{\mathcal{U}}\big(u- \eta (\underbrace{\nabla f(u) + \Pi_u^{\sf T} \nabla g(\Pi_u u +\Pi_w w + \Tilde{v}))}_{\define F_{w}(u,\tilde{v})}\big)-u
    \end{aligned}
\end{equation}
where $\tilde{v}\define \mathcal{C}\tilde{\zeta}=\begin{bmatrix}
    C& -\Pi_w 
\end{bmatrix}\tilde{\zeta}$.
As shown in Figure \ref{Fig:Cascaded}, the closed-loop system \eqref{Eq:LTI}, \eqref{Eq:EstErrorDynamics}, \eqref{Eq:EstErrorController} is a cascade, with the estimator error dynamics \eqref{Eq:EstErrorDynamics} driving the gradient flow controller \eqref{Eq:EstErrorController}, which in turn drives the plant \eqref{Eq:LTI}.

\begin{figure}[ht!]
    \centering
    \input {Figures/fig2.tex}
    \caption{Block diagram of closed-loop system in error coordinates, where $\mathcal{H}_{w}(u,\tilde{v}) = P_{\mathcal{U}}(u - \eta F_w(u,\Tilde{v})) - u$.}
    \label{Fig:Cascaded}
\end{figure}
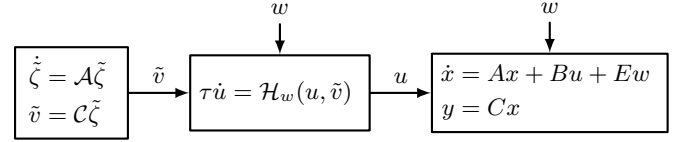

Since $\mathcal{A}$ is Hurwitz, the unique equilibrium of \eqref{Eq:EstErrorDynamics} is $\tilde{\zeta} = (\tilde{x},\tilde{w}) = (0,0)$, with corresponding equilibrium value of $\tilde{v}$ also being zero. Equilibria of \eqref{Eq:EstErrorController} are then determined by
\begin{equation}\label{Eq:EquilibriumController}
\small{
0 = P_{\mathcal{U}}\big(\bar u-F_{w}(\bar{u},0)\big)-\bar u. 
}
\end{equation}
Under Assumption \ref{Ass:Cost}, $u \mapsto F_{w}(u,0)$ is $\mu_f$-strongly monotone and $(\ell_f + \ell_g \|\Pi_u\|_2^2)$-Lipschitz, and thus by Proposition \eqref{Prop:ControllerISS}, \eqref{Eq:EquilibriumController} possesses a unique solution in $\mathcal{U}$. 
Moreover, under Assumption \ref{Ass:Cost}, \eqref{Eq:EquilibriumController} is precisely the (necessary and sufficient) KKT condition for optimality of \eqref{optimization_simple}, and thus the global optimizer $\bar{u}^{\star} \in \mathcal{U}$ is the unique solution of \eqref{Eq:EquilibriumController}. Finally, since $A$ is Hurwitz, $\bar{x} = -A^{-1}B\bar{u}^{\star} - A^{-1}Ew$ is the unique equilibrium of the plant with corresponding output $\bar{y}^{\star}$. This shows statement (i). Since \eqref{Eq:EstErrorDynamics} and \eqref{Eq:LTI} are exponentially stable linear systems, to show global exponential stability of the three-system cascade, it suffices (e.g., \cite[Lemma 4.7]{HKK:02}) to show that \eqref{Eq:EstErrorController} is (exponentially) input-to-state stable with respect to its equilibrium point $\bar{u}^{\star}$, which now follows from Proposition \ref{Prop:ControllerISS}. This shows (ii) and completes the proof.
\end{pfof}

Theorem \ref{Thm:EEFBO} requires no limitations on the tuning parameter $\tau$, which can be arbitrarily small. As shown in Figure \ref{Fig:Cascaded}, the closed-loop system of \eqref{Eq:LTI} and \eqref{Eq:FBOEstimator} has the form of a three-system cascade when expressed in certain error coordinates. An important implication is that \emph{the slowest convergence rate among the estimator, controller, and plant will determine the closed-loop convergence rate}. Therefore, if the estimator and the controller are designed to be sufficiently fast, the convergence rate of the closed-loop system will equal the convergence rate of the open-loop plant which is a significant improvement over the baseline FBO design \eqref{Eq:GradientController}.

\subsection{Model-Information vs. Performance Trade-Offs in FBO}


The above analysis highlights a trade-off between the amount of model information used and the tuning limitations (i.e., the achievable control performance) of the design. On one end of the spectrum, the traditional FBO approach \eqref{Eq:GradientController} relies only on the equilibrium mapping (DC gain) of the LTI system; this simplifies design, but imposes limitations on closed-loop tuning, leading to degraded transient performance. 
In contrast, the estimator-enhanced FBO scheme \eqref{Eq:FBOEstimator} exploits more comprehensive model information to generate real-time estimates of the plant output and the unmeasured disturbance. By incorporating this additional information, the controller can more accurately predict and compensate for transient deviations, eliminating the tuning restrictions in traditional FBO. and enabling improved transient performance.

Recognizing this trade-off motivates the exploration of intermediate FBO designs that incorporate partial dynamic model information, such as reduced-order models capturing dominant dynamics. In Section \ref{Sec:ReducedEstFBO}, we explore this by designing an estimator based on a reduced-order plant model tailored for two-timescale systems. This design illustrates one possible compromise between model information used and improvement in FBO control performance. 



\section{Estimator-Based FBO for Two-Timescale Systems}\label{Sec:ReducedEstFBO}

\subsection{Estimator Design Based on Reduced Plant Model}
\label{Sec:ReducedEstFBOMotivation}

In this section, we assume the plant is described by the two-timescale LTI model \cite{PVK-HKK-JO:99}
\begin{equation}\label{Eq:LTISingPert}
\begin{aligned}
    \begin{bmatrix}
        \dot{x}_1\\
        \varepsilon\dot{x}_2
    \end{bmatrix}&=\begin{bmatrix}
        A_{11} & A_{12} \\
        A_{21} & A_{22}
    \end{bmatrix}\begin{bmatrix}
        x_1\\
        x_2
    \end{bmatrix}+\begin{bmatrix}
        B_1\\
     B_2
    \end{bmatrix}u+\begin{bmatrix}
        E_1\\
        E_2
    \end{bmatrix}w\\
    y &= \begin{bmatrix}
        C_1 & C_2
    \end{bmatrix}\begin{bmatrix}
        x_1\\
        x_2
    \end{bmatrix},
\end{aligned}
\end{equation}
where $\varepsilon>0$, with slow state $x_1(t) \in \real^{n_1}$, fast state $x_2(t) \in \real^{n_2}$, control input $u(t) \in \real^m$, measured output $y(t) \in \mathbb{R}^p$, and where $w \in \mathbb{R}^q$ is a \textit{constant} disturbance.
We again assume that the open-loop plant is stable, i.e., $\mathrm{blkdiag}(I_{n_1},\varepsilon I_{n_2})^{-1}A$ is Hurwitz, and additionally that $A_{22}$ is Hurwitz.

For the two-timescale system \eqref{Eq:LTISingPert}, a standard reduced model can be derived which captures the dynamics of the slower timescale of the system. In particular, when $\varepsilon$ is small, the fast states converge to the quasi steady-state values 
\begin{equation}\label{Eq:x2Quasi}
\bar x_2^{\rm r}=-A_{22}^{-1}A_{21}x_1- A_{22}^{-1}B_2 u - A_{22}^{-1}E_2 w,
\end{equation}
which upon substitution into the $x_1$ dynamics lead to the \emph{reduced model}
\begin{equation}\label{Eq:ReducedModel}
\begin{aligned}
\dot{x}_1^{\rm r} &= A_0x_1^{\rm r} + B_0u + E_0 w\\
y^{\rm r} &= C_0x_1^{\rm r} + D_0u +Q_0 w,
\end{aligned}
\end{equation}
where
\begin{equation}
\label{eq:reduced_matrices}
\begin{aligned}
A_0 &\define A_{11}-A_{12}A_{22}^{-1}A_{21}, &\quad  B_0 &\define B_1,\\[-0.3em]
C_0 &\define C_1-C_2A_{22}^{-1}A_{21},       &\quad  D_0 &\define -C_2A_{22}^{-1}B_2,\\[-0.3em]
E_0 &\define E_1-A_{12}A_{22}^{-1}E_2,       &\quad  Q_0 &\define -C_2A_{22}^{-1}E_2.
\end{aligned}
\end{equation}
Since $\mathrm{blkdiag}(I_{n_1},\varepsilon I_{n_2})^{-1}A$ and \(A_{22}\) are Hurwitz, Tikhonov's Theorem implies that $A_0$ is Hurwitz \cite[Theorem 11.1]{HKK:02}. Furthermore, it is straightforward to establish that the DC gain matrices $\Pi_{u}$ and $\Pi_{w}$ of \eqref{Eq:LTISingPert} can be equivalently expressed using the reduced model as
\begin{equation}\label{Eq:LTISingPertDCGain}
\Pi_{u} = D_0-C_0A_0^{-1}B_0, \qquad \Pi_{w} = Q_0-C_0A_0^{-1}E_0.
\end{equation}

We now assume that \emph{only the reduced model \eqref{Eq:ReducedModel}} is available for estimator design, rather than the full system model \eqref{Eq:LTISingPert}. Following the approach fo Section \ref{Sec:EstFBO}, we consider the augmented model
\begin{equation}\label{Eq:ReducedAugmented}
\begin{aligned}
\begin{bmatrix}
\dot{x}_1^{\rm r}\\
\dot{w}
\end{bmatrix} &= \underbrace{\begin{bmatrix}A_0 & E_0 \\ 
\vzeros & \vzeros
\end{bmatrix}}_{\define \mathbf{A}_{\rm aug0}}\begin{bmatrix}x_1^{\rm r} \\ w\end{bmatrix} + \underbrace{\begin{bmatrix}B_0 \\ \vzeros \end{bmatrix}}_{\define \mathbf{B}_{\rm aug0}}u\\
y^{\rm r} &= \underbrace{\begin{bmatrix}C_0 & Q_0\end{bmatrix}}_{\define \mathbf{C}_{\rm aug0}}\begin{bmatrix}x_1^{\rm r} \\ w\end{bmatrix}+ D_0 u,
\end{aligned}
\end{equation}
and mirroring Assumption \ref{Ass:NonRes}, we require a non-resonance condition. We record this and the previous stability assumptions for convenience.

\begin{assumption}[\bf Two-Timescale System Assumptions]\label{Ass:TwoTime}
The matrices $\mathrm{blkdiag}(I_{n_1},\varepsilon I_{n_2})^{-1}A$ and $A_{22}$ are Hurwitz, and 
\begin{equation}\label{Eq:NonResSingPert}
\mathrm{rank} \begin{bmatrix}A_0 & E_0\\ C_0 & 0\end{bmatrix} = n_1 + q.
\end{equation}
\end{assumption}

\smallskip

The proposed EE-FBO design is


\begin{subequations}\label{Eq:ReducedFBOEstimator}
    \begin{align}
    \label{Eq:ReducedFBOEstimator-1}
    \begin{bmatrix}
    \dot{\hat{x}}_1^{\rm r}\\
    \dot{\hat{w}}
    \end{bmatrix} &= \mathbf{A}_{\rm aug0}\begin{bmatrix}\hat{x}_1^{\rm r} \\ \hat{w}\end{bmatrix} + \mathbf{B}_{\rm aug0}u - \mathbf{L}_0(y - \hat{y}^{\rm r})\\
    \label{Eq:ReducedFBOEstimator-3}
    \hat{y}^{\rm r} &= \mathbf{C}_{\rm aug0}\begin{bmatrix}\hat{x}_1^r \\ \hat{w}\end{bmatrix}+D_0u\\
    \label{Eq:ReducedFBOEstimator-4}
    \hat{\bar{y}}&=\Pi_u u+\Pi_w \hat{w}\\
    \label{Eq:ReducedFBOEstimator-5}
    \tau \dot{u}&=P_{\mathcal{U}}\Big(u-\eta \Phi (u,y - (\hat y^{\rm r} - \hat{\bar y})\big)\Big)-u,
    \end{align}
\end{subequations}
where $\mathbf{L}_0 \in \real^{(n_1+ q) \times p}$ is an estimator gain to be designed, $\eta, \tau > 0$, and $\Phi$ is as in \eqref{Eq:Phi}.

\subsection{Stability Analysis}\label{Sec:ReducedStabilityAnalysis}

We can now state the second main result, concerning stability of the closed-loop system \eqref{Eq:LTISingPert},\eqref{Eq:ReducedFBOEstimator}.

\smallskip

\begin{theorem}[\bf Unconditional Stability of EE-FBO Based on Reduced Model]\label{Thm:ReducedEEFBO}
Under Assumptions \ref{Ass:Cost} and \ref{Ass:TwoTime}, select $\eta \in (0,2\mu_{f}/(\ell_f + \ell_g \|\Pi_u\|_{2}^2)^2)$, and select $\mathbf L_0$ such that $\mathbf A_{\rm aug0} + \mathbf L_0 \mathbf C_{\rm aug0}$ is Hurwitz. Then (i) for each $w \in \real^{q}$, the closed-loop system \eqref{Eq:LTISingPert}, \eqref{Eq:ReducedFBOEstimator} possesses a unique equilibrium $(x_1,x_2,\hat{x}_{1}^{\rm r},\hat{w},u)=(\bar{x}_1,\bar{x}_2,\bar{x}_1,w,\bar{u}^{\star})$ with corresponding output $\bar{y}^{\star}$, and (ii) there exists $\varepsilon^* > 0$ such that if $\varepsilon \in (0,\varepsilon^*)$, the equilibrium point is globally exponentially stable \underline{for all} $\tau > 0$.
\end{theorem}

\smallskip

Theorem \ref{Thm:ReducedEEFBO} essentially states that as long as there is ``enough'' timescale separation in the plant \eqref{Eq:LTISingPert}, the unconditional stability result of Theorem \ref{Thm:EEFBO} persists. We emphasize that this is a timescale separation \emph{internal} to the two-timescale plant, and \textbf{not} a timescale separation between the FBO design and the slow dynamics of the plant.

\smallskip

\begin{pfof}{Theorem \ref{Thm:ReducedEEFBO}}
%
%
%
The arguments regarding existence/uniqueness of the equilibrium are as in the proof of Theorem \ref{Thm:EEFBO}, and are omitted for brevity. Under Assumption \ref{Ass:TwoTime}, it is straightforward to show that \eqref{Eq:ReducedAugmented} is detectable, and thus $\mathbf{L}_0 = \mathrm{col}(L_1,L_2)$ can be selected such that $\mathcal{A}\define \mathbf{A}_{\rm aug0} + \mathbf{L}_0\mathbf{C}_{\rm aug0}$ is Hurwitz. Writing the plant \eqref{Eq:LTISingPert} in new coordinate defined as $x_2 \mapsto e_2 = x_2 - \bar{x}_2^{\rm r}$, where $\bar{x}_{2}^{\rm r}$ is defined in \eqref{Eq:x2Quasi}, we obtain
\begin{equation}\label{Eq:LTI-TT-Transformed}
\begin{aligned}
\dot{x}_1 &= A_0 x_1 + A_{12} e_2 + B_0 u + E_0 w, \\
\varepsilon \dot{e}_2 &= A_{22} e_2 + \varepsilon (M_1 x_1 + M_2 e_2) + \varepsilon M_3 u + \varepsilon M_4 w, \\
y &= C_0 x_1 + C_2 e_2 + D_0 u+ Q_0 w.
\end{aligned}
\end{equation}
where 
\begin{align*}
M_1 &\define A_{22}^{-1}A_{21}A_0, \quad  M_2 \define A_{22}^{-1}A_{21}A_{12} ,\\
M_3 &\define A_{22}^{-1}A_{21}B_0, \quad M_4\define A_{22}^{-1}A_{21}E_0. 
\end{align*}
and $(A_0,B_0,C_0,E_0,D_0,Q_0)$ are the reduced model parameters from \eqref{eq:reduced_matrices}. Using $e_2$ and defining the estimation error $\tilde{\zeta} = (\tilde{x}_1, \tilde{w}) \define (x_1 - \hat{x}_1, w - \hat{w})$, the estimator dynamics \eqref{Eq:ReducedFBOEstimator-1}--\eqref{Eq:ReducedFBOEstimator-4} become
\begin{equation}\label{Eq: SingPertbEstError}
\small
\begin{aligned}
\dot{\tilde{\zeta}} &=\mathcal{A}\tilde{\zeta} + \mathcal{G}e_2
\\
\tilde y
&= [\,C_0 \;\; D_0\,]\,
  \tilde{\zeta}
  + C_2\,e_2
\\
\hat{\bar{y}}
&= \Pi_u\,u + \Pi_w\,\hat w
= \Pi_u\,u + \Pi_w\,(w - \tilde w),
\end{aligned}\,\,
\begin{aligned}
\mathcal{G} \define \begin{bmatrix}
    A_{12}-L_1C_2\\
    -L_2C_2
  \end{bmatrix}
\end{aligned}
\end{equation}
and the controller \eqref{Eq:ReducedFBOEstimator-5} accordingly becomes
\begin{equation}\label{Eq: SingPertController}
\begin{aligned}
\tau \dot u
&= P_{\mathcal{U}}(u - \eta F_{w}(u,\tilde{v},e_2)) - u
\end{aligned}
\end{equation}
where
\[
\begin{aligned}
F_{w}(u,\tilde{v},e_2) = \nabla f(u)
   +\Pi_u^{\T}\,\nabla g\bigl(\bar y +\tilde v +C_2 e_2\bigr)
\tilde v &= \mathcal{C}\tilde\zeta, 
   \end{aligned}
\]
with $\bar y= \Pi_u u+\Pi_w w$ and where $\mathcal{C} \define \begin{bmatrix}C_{0} & D_{0}-\Pi_{w}\end{bmatrix}$. The closed-loop system \eqref{Eq:LTI-TT-Transformed}--\eqref{Eq: SingPertController} now has the structure shown in Figure \ref{fig:InterconnectedSys}, and we proceed via small-gain arguments. 
\begin{figure}[ht!]
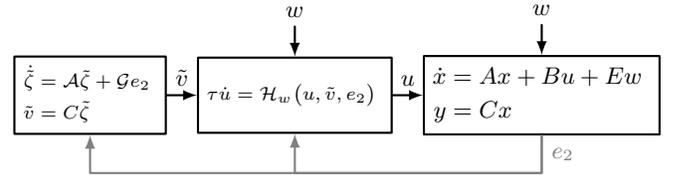
 
    \centering 
    \include{Figures/fig4} 
    \caption{Block diagram of the interconnected estimator, controller, and plant, where $\mathcal{H}_{w}(u,\tilde{v},e_2) = P_{\mathcal{U}}(u - \eta F_{w}(u,\tilde{v},e_2)) - u$.} 
    \label{fig:InterconnectedSys} 
\end{figure} 

Since $\mathcal{A}$ is Hurwitz, the estimator \eqref{Eq: SingPertbEstError} has finite $\mathsf{L}_2$-gain from input $e_2$ to output $\tilde{v}$ which we denote by $\gamma_{e_2\to \tilde{v}}$. The assumptions of  Proposition~\ref{Prop:ControllerISS} hold with $\mu = \mu_f$, $L = \ell_f + \ell_g \|\Pi_u\|_2^2$, $L_v^{\prime} = \ell_g \|\Pi_u\|^2$, $L_e^{\prime} = \ell_g \|\Pi_u\|_2^2 \|\mathcal{C}\|_2$, and the controller block \eqref{Eq: SingPertController} has finite $\mathsf{L}_2$-gains $\gamma_{\tilde v\to u}, \gamma_{e_2\to u}$ as defined in \eqref{Eq:ControllerL2Gains}. It follows that the cascade of the estimator and controller has a $\mathsf{L}_2$-gain from $e_2$ to $u - \bar{u}^*$ bounded by
\(
\gamma \define \gamma_{e_2\to u} + \gamma_{e_2\to \tilde v}\gamma_{\,\tilde v\to u}.
\)
Returning to the plant \eqref{Eq:LTI-TT-Transformed}, by Assumption \ref{Ass:TwoTime} there exist \(P_0,P_{22}\succ0\) and \(\alpha_0,\alpha_{22} > 0\) such that
\begin{align*}
P_0 A_0 + A_0^\T P_0 &\preceq -\alpha_0 I_{n_1} \\
P_{22} A_{22} + A_{22}^\T P_{22}  &\preceq -\alpha_{22} I_{n_2}.
\end{align*}
With $\tilde u \define u-\bar u^*$ and the composite storage function $V(x_1,e_2) \define x_1^\T P_0x_1+ e_2^\T P_{22}e_2$, tedious but routine calculations show that the $\mathsf{L}_2$-gain of \eqref{Eq:LTI-TT-Transformed} from $\tilde{u}$ to $e_2$ is upper bounded by
\[
\gamma_{\rm p} \define \frac{\beta}{\tfrac{\alpha_{22}}{\varepsilon} - \alpha} > 0,
\]
where $\alpha,\beta > 0$ depend on the Lyapunov variables and plant parameters, but are independent of $\varepsilon$. In particular, note that $\gamma_{\rm p}$ is $O(\varepsilon)$ as $\varepsilon \to 0$.  Global exponential stability of the equilibrium now follows from the small-gain theorem \cite[Theorem~5.6]{HKK:02} if $\gamma \gamma_{\rm p} < 1$, which yields the condition
\[
\varepsilon < \varepsilon^* \define \frac{\alpha_{22}}{\beta \gamma^{2}+\alpha}
\]
and completes the proof of (ii).
\end{pfof}

\section{Application to Fast Frequency Control of Power Systems using IBRs}\label{Sec:Example}

We demonstrate and compare the performance and stability of three control approaches \textemdash{} traditional FBO, EE-FBO with the full plant model, and EE-FBO using a reduced-order plant model \textemdash{} on an application problem of power system frequency control using inverter-based resources (IBRs). The set-up follows that in \cite{EE-JWSP-EF-MP-AH-LZ:23f}. The plant is a small-signal lumped mechanical frequency dynamics model of a single-area transmission system with $N$ fast acting dispatchable IBRs, and a constant unknown power lumped imbalance, given by
\begin{subequations}\label{Eq:SFR}
\begin{align}
    2H\Delta{\dot{\omega}} &= - D\Delta \omega + \Delta P_{\mathrm{m}} - \Delta P_{\mathrm{u}} + \sum_{i=1}^{N}\nolimits \Delta P_{\mathrm{ibr},i}\\
    T_{\rm R}\Delta{\dot{P}_{\mathrm{m}}} &= - \Delta P_{\mathrm{m}} - R_{\rm g}^{-1}(\Delta{\omega} + T_{\mathrm{R}}F_{\mathrm{H}}\Delta{\dot{\omega}})\\
    \label{Eq:SFR-IBR}
    \tau_i \Delta\dot{P}_{\mathrm {ibr},i} &= -\Delta P_{\mathrm{ibr},i} + u_i, \qquad i \in \{1,\ldots,N\},
\end{align}
\end{subequations}
where $\Delta \omega$ [p.u.] is the frequency deviation, $\Delta P_{\rm m} $ [p.u.] is mechanical power change, $\Delta P_{\rm u}$ is the unmeasured constant disturbance, $\Delta P_{\mathrm{ibr,i}}$ is the IBR power change, and $u_i$ is the IBR power change set-point. The parameter $2H= 26.3083$ [s] is the inertia constant, $D =0 $ [p.u.] is the load damping, $T_{\rm R}= 10$ [s] is the reheat time constant, $R_{\rm g} =0.05$ [p.u.] is the generator droop constant, $F_{\rm H}= 0.64$ is the fraction of total power generated by the pressure turbine. We consider $N = 2$, with IBR time constants and $\tau_1,\tau_2=0.3$ [s]. Unless stated otherwise, all quantities are on the system base $S_{\text{base}} = 567.5~\text{MW}$ and frequency base $f_{\text{base}} = 60~\text{Hz}$ ($\omega_{\text{base}} = 2\pi f_{\text{base}}$).

\begin{figure*}[!ht]
  \centering
  \begin{subfigure}{0.24\textwidth}
    \includegraphics[width=\linewidth]{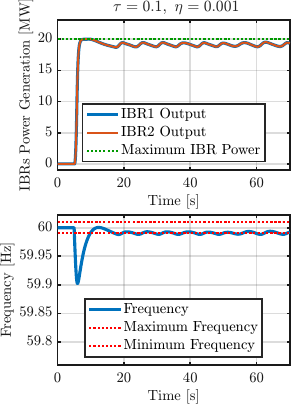}
    \caption{Unstable original FBO}
    \label{fig:FBO_orig_unstable}
  \end{subfigure}
  \begin{subfigure}{0.24\textwidth}
    \includegraphics[width=\linewidth]{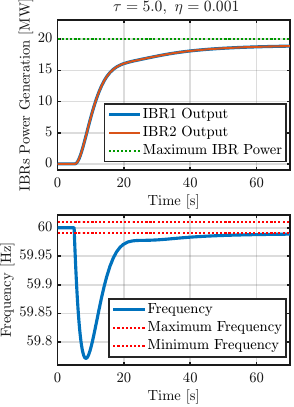}
    \caption{Stable original FBO}
    \label{fig:FBO_orig}
  \end{subfigure}\hfill
  \begin{subfigure}{0.24\textwidth}
    \includegraphics[width=\linewidth]{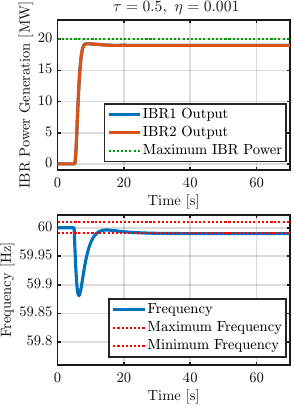}
    \caption{EE-FBO based on reduced model}
    \label{fig:FBO_red}
  \end{subfigure}\hfill
  \begin{subfigure}{0.24\textwidth}
    \includegraphics[width=\linewidth]{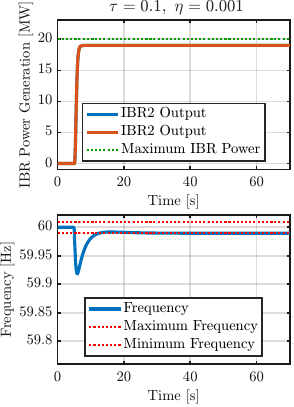}
    \caption{EE-FBO based on full model}
    \label{fig:FBO_full}
  \end{subfigure}\hfill
  \caption{Response power system under $40$MW load change under FBO and EE-FBO methods.}
  \label{fig:FBO_compare}
\end{figure*}



The control objective is to maintain the frequency deviation $y = \Delta \omega$ within prescribed bounds $[\underline{y},\overline{y}] = [-0.01,0.01]$ while minimizing a cost of IBR power usage and satisfying the power command limits $(u_1,u_2) \in \mathcal{U}= [\,-20\,\text{MW},20\,\text{MW}]^2$. This is encoded through the objective functions
\[
f(u) = \tfrac{1}{2}\|u\|^2, \quad g(y) = \tfrac{1}{2}10^{5} \max\{0, \underline{y} - y, y - \bar{y}\}^2.
\]
The resulting function $\Phi$ from \eqref{Eq:Phi} used in all controllers takes the form $\Phi(u,y) = u  + 10^{5}S_{\underline{y},\bar{y}}(y)$, where
\[
S_{\underline{y},\bar{y}}(y) = \begin{cases}
y - \underline{y}, & y < \underline{y} \\
0, & \underline{y} \leq y \leq \bar{y} \\
y - \bar{y}, & y > \bar{y}.
\end{cases}
\]

For the EE-FBO methods, the required estimators are designed using the standard linear-quadratic estimation (LQE) method. In particular, for the full-model EE-FBO we take 
\[
Q=\operatorname{diag}(10^{-2},10^{-2},10^{2},10^{2},10^{6}), \qquad R=1,
\]
where the states correspond to those in \eqref{Eq:SFR} plus the disturbance $\Delta P_{\rm u}$. For the reduced-model EE-FBO, the IBR dynamics \eqref{Eq:SFR-IBR} are fast and are eliminated as described in Section \ref{Sec:ReducedEstFBO}; the LQE estimator for the resulting reduced model with states $(\Delta \omega,\Delta_{\rm m},\Delta P_{\rm u})$ is designed using weights
\[
Q=\operatorname{diag}(10^{-2},10^{-2},10^{6}), \qquad R=1.
\]

We simulate the closed-loop response under the three methods across a range of tuning parameters \(\tau\) with a load change of $\Delta P_u = 40 \;\text{[MW]}$ occurring at $t=5\; [\text{s}]$. As shown in Figure \ref{fig:FBO_orig_unstable}, with small $\tau$ in the traditional FBO method, the response is unstable and shows sustained oscillations. To eliminate the oscillations, $\tau$ must be increased (Figure \ref{fig:FBO_orig}), leading to slow regulation of the frequency and a low frequency nadir. In contrast, both the reduced-model EE-FBO (Figure \ref{fig:FBO_red}) and the full-model EE-FBO (Figure \ref{fig:FBO_full}) maintain stability even at small \(\tau\), allowing for a significantly improved transient response. In particular, the reduced-model EE-FBO shows only slightly degraded performance compared to the full model EE-FBO, thus achieving a reasonable trade-off between performance and model information used. These results confirm that EE-FBO methods can remove the traditional speed limitations present in FBO, and improve closed-loop performance.



\section{Conclusion}\label{Sec:Conclusion}

We have presented an approach which addresses the timescale separation limitation of traditional FBO design. The key idea is to design an estimator based on model information, use it to generate an estimate of the systems’s steady-state output, and integrate this estimate into the traditional FBO controller. We established rigorous stability results for (i) the case of estimator design based on a full dynamic model, and (ii) the case of estimator design based on an approximate dynamic model when the plant itself is singularly perturbed. In either case, the achievable closed-loop convergence rate is limited only by the dominant eigenvalue of the plant and the designed estimator. A frequency control case study showed significant performance gains over FBO, with the reduced-model design achieving performance close to that of the full-model design. Our work points towards a spectrum of possible FBO designs, which trade off model information against achievable control performance. Future work will consider robust and data-driven estimator designs that further reduce reliance on explicit plant models, extensions to nonlinear systems, and more comprehensive application of the results to problems in energy systems and robotics.

\bibliographystyle{IEEEtran}
\bibliography{brevalias, JWSP, Main, ref}

\appendices
\section{Proofs}
\label{Sec:Proof}

\begin{pfof}{Proposition \ref{Prop:ControllerISS}} 
Let $T_{v,e}(u) = P_{\mathcal{U}}(u-\eta \pi(u,v,e))$. The dynamics can be written as
\[
\tau\dot{u} = T_{v,e}(u) - u = T_{0,0}(u) - u + [T_{v,e}(u) - T_{0,0}(u)]. 
\]
%
%
By standard arguments \cite[Theorem 12.1.2]{FF-JSP:03b}, for $\eta < 2\mu/L^2$ the map $T_0(u)$ is a contraction map on the complete space $(\mathcal{U},\|\cdot\|_2)$, and thus
\[
\|T_{0,0}(u) - T_{0,0}(u^{\prime})\|_2 \leq \rho \|u-u^{\prime}\|_2
\]
for any $u,u^{\prime}$, where $\rho = \sqrt{1- 2\eta\mu + \eta^2 L^2} \in (0,1)$. By Banach’s fixed-point theorem \cite[Theorem B.1]{HKK:02}, there exists a unique $\bar u^*\in\mathcal U$ such that
\[
\bar u^* \;=\; T_{0,0}(\bar u^*) \;=\; P_{\mathcal U}\!\big(\bar u^*-\eta\,\pi(\bar u^*,0,0)\big).
\]
This $\bar u^*$ is thus the unique equilibrium of $\dot u = T_0(u)-u$. With $V(u) = \tfrac{1}{2}\|u-\bar{u}^*\|_2^2$, routine calculations show that
\[
\begin{aligned}
\dot{V} &= -(1-\rho)\|u-\bar{u}^*\|_2^2 + (u-\bar{u}^*)^{\sf T}(T_{v,e}(u) - T_0(u))\\
&\leq -2(1-\rho)V + \|u-\bar{u}^*\|_2 \|T_{v,e}(u) - T_{0,0}(u)\|_2
\end{aligned}
\]
Using non-expansiveness of projections, we have
\[
\begin{aligned}
\small
\|T_{v,e}(u) - T_{0,0}(u)\|_2 &\leq \|u - \eta \pi(u, v,e) - (u - \eta \pi(u, 0))\|_2\\
&\leq \eta L_v^{\prime}\|v\|_2+ \eta L_e^{\prime}\|e\|_2
\end{aligned}
\]
so
\[
\begin{aligned}
\dot{V} &\leq -2(1-\rho)V + \eta L_v^{\prime}\|u-\bar{u}^*\|_2 \|v\|_2+ \eta L_e^{\prime}\|u-\bar{u}^*\|_2 \|e\|_2\\
&\leq -2(1-\rho)V \\
 &+ \left(c_e V + \tfrac{\eta^2}{2c_e} (L_v^{\prime})^2\|v\|_2^2\right) +  \left(c_v V + \tfrac{\eta^2}{2c_v} (L_e^{\prime})^2\|e\|_2^2\right)\\
\end{aligned}
\]
for any $c_v,c_e > 0$. Collecting terms, we find that
\begin{equation}\label{Eq:ControllerISSLyap}
\dot{V} \leq -[2(1-\rho)-c_v-c_e]V + \tfrac{\eta^2 {L_v^{\prime}}^2}{2c_v}\|v\|_2^2 + \tfrac{\eta^2 {L_e^{\prime}}^2}{2c_e}\|e\|_2^2.
\end{equation}

We must choose $c_v+c_e < 2(1-\rho)$. Therefore, choosing $c_e=c_v = (1-\rho)/2$ and by comparison \cite[Lemma 3.4]{HKK:02}, the inequality now implies the ISS bound
\[
\begin{aligned}
\small
\|u(t)-\bar{u}^*\|_2 \leq& e^{-\tfrac{1}{2}(1-\rho)t}\|u(0)-\bar{u}^*\|_2  \\
&+ \frac{\eta L_v^{\prime}}{1-\rho} \sup_{s \geq t}\|v(s)\|_2+ \frac{\eta L_e^{\prime}}{1-\rho} \sup_{s \geq t}\|e(s)\|_2,
\end{aligned}
\]
which shows the result.
\end{pfof}

\end{document}

%% file: confpaperdefs.tex
\usepackage{etex}
\usepackage[vlined,ruled]{algorithm2e}
\usepackage[dvipsnames]{xcolor}
\usepackage{pgfpages}
\usepackage[cmex10]{amsmath}		
\usepackage{amssymb, mathrsfs}
\usepackage{cite}
\usepackage{url}
\usepackage{subcaption} 

\usepackage{dsfont}
\usepackage{siunitx}
\usepackage{verbatim}									
\usepackage{arydshln}
\usepackage{mathtools}									
\usepackage{siunitx}									
\usepackage{mdframed}

\usepackage{graphicx}
\usepackage{epstopdf}

\usepackage[font=small]{caption}
\usepackage{subcaption}
\usepackage[activate={true,nocompatibility},final,tracking=true,kerning=true,spacing=true,factor=1100,stretch=10,shrink=10]{microtype}

\usepackage{array}
\usepackage{multirow}
\usepackage{enumerate}
\usepackage{booktabs}

\usepackage{graphicx}
\usepackage{subcaption}   
\graphicspath{{Figures/}}

\graphicspath{{./fig/}}

\usepackage{pgfplots}
\pgfplotsset{width=8cm,compat=newest}
\newcommand*{\damping}{0.006}%
\newcommand*{\freq}{25}%
\pgfmathsetmacro{\freqd}{sqrt(1-(\damping)^2)*\freq}%

\pgfplotsset{
    standard/.style={
    axis x line=middle,
    axis y line=middle,
    enlarge x limits=0.15,
	enlarge y limits=0.15,
	every axis plot post/.style={mark options={fill=black}},
	}
}

\pgfplotsset{%
    ,compat=1.12
    ,every axis x label/.style={at={(current axis.right of origin)},anchor=north west}
    ,every axis y label/.style={at={(current axis.above origin)},anchor=north east}
    }

\usepackage{tikz}
\usetikzlibrary{arrows}
\usetikzlibrary{decorations.pathmorphing}
\usetikzlibrary{decorations.markings}
\usetikzlibrary{snakes}
\usetikzlibrary{matrix}
\usetikzlibrary{calc}
\usetikzlibrary{patterns}
\usetikzlibrary{positioning}
\usetikzlibrary{shapes}
\usetikzlibrary{shapes.symbols}
\usetikzlibrary{shapes.geometric}
\usetikzlibrary{shadows.blur}
\usetikzlibrary{shapes.arrows}
\usetikzlibrary{shapes.multipart}
\usetikzlibrary{shapes.callouts}
\usetikzlibrary{shapes.misc}

\tikzstyle{every node}=[font=\small]
\tikzstyle{every path}=[line width=0.8pt,line cap=round,line join=round]

\usepackage[american,smartlabels]{circuitikz}
\ctikzset{bipoles/thickness=1}
\ctikzset{bipoles/length=0.8cm}
\ctikzset{bipoles/diode/height=.375}
\ctikzset{bipoles/diode/width=.3}
\ctikzset{tripoles/thyristor/height=.8}
\ctikzset{tripoles/thyristor/width=1}
\ctikzset{bipoles/vsourceam/height/.initial=.7}
\ctikzset{bipoles/vsourceam/width/.initial=.7}

\newcommand{\real}{\mathbb{R}}

\DeclareMathOperator*{\minimize}{minimize} 									

\newcommand{\vect}[1]{\mathbbold{#1}}

\newcommand{\vzeros}[1][]{\vect{0}_{#1}}
\DeclareSymbolFont{bbold}{U}{bbold}{m}{n}
\DeclareSymbolFontAlphabet{\mathbbold}{bbold}

\newcommand{\map}[3]{#1: #2 \rightarrow #3}

\usepackage{soul}

\newcommand{\T}{\mathsf{T}}

\DeclareMathOperator{\subto}{subject~to}

\newcommand\oprocendsymbol{\hbox{$\square$}}
\newcommand\oprocend{\relax\ifmmode\else\unskip\hfill\fi\oprocendsymbol}

\newtheorem{theorem}{Theorem}[section]

\newtheorem{proposition}[theorem]{Proposition}

\newtheorem{assumption}{Assumption}[section]

\newenvironment{pfof}[1]{\vspace{1ex}\noindent{\itshape Proof of
    #1:}\hspace{0.5em}} {\hfill\oprocend\vspace{1ex}}

\newcommand{\define}{\coloneqq}

\newif\ifforstudents

\pgfdeclarelayer{background}
\pgfsetlayers{background,main}

%% file: Figures/fig1.tex
    \begin{tikzpicture}[
        block/.style = {
          rectangle, draw=black, fill=white,
          thick, 
          minimum width=2cm, minimum height=1cm, align=center
        },
        int/.style = {
          rectangle, draw=black, fill=white,
          line width=1pt, 
          minimum width=1cm, minimum height=2cm, align=center
        },
        >=latex
      ]
      \node[block]                     (plant) {
        $\begin{aligned}
              \dot{x}&=Ax+Bu+Ew\\
              y &= Cx
          \end{aligned}$};
      \node[block, below=1cm of plant] (est)   {Estimator};
      \node[block, below=0.5cm of est]   (opt)   {
        $\tau \dot{u} = \mathcal{P}_{\mathcal{U}}\Big(u- \eta \Phi(u,y-(\hat{y}-\hat{\bar y}))\Big)-u$
      };

      \node[above=0.5cm of plant] (w) {$w$};
      \draw[->, thick] (w) -- (plant.north);

      \draw[->, thick]
        (plant.east) -- ++(2.5cm,0) node[black,midway, above] {$y$} coordinate (yout);

      \draw[-, thick]
        ([xshift=2cm]plant.east) -- ++ (0,-1cm) coordinate (y1);
       \draw[-, thick]
        (y1) -- ++(-2cm,0) -| ([xshift=-1cm,yshift=0.1cm]est.west) coordinate (y2) ;
        \draw[->, thick]
        (y2) -- ([yshift=0.1cm]est.west);
        \draw[->, thick]
        (y1) -- ++(0,-2cm) |- ([yshift=-0.3cm]opt.east);

      \draw[->, thick]
        (est.east) -- ++(2.3cm,0) node[black,midway, above] {$\hat y, \hat{\bar{y}}$} |- ([yshift=0.1cm]opt.east);

      \draw[->, thick]
        (opt.west) -- ++(-0.4cm,0) node[black,midway,above] {$u$} |- (plant.west);
      \draw[->, thick]
        (opt.west) -- ++(-0.4cm,0) |- ([yshift=-0.2cm]est.west);
    \end{tikzpicture}

%% file: Figures/fig2.tex
\begin{tikzpicture}[
        block/.style = {
          rectangle, draw=black, fill=white,
          thick, 
          minimum width=1.5cm, minimum height=1cm, align=center
        },
        int/.style = {
          rectangle, draw=black, fill=white,
          thick, 
          minimum width=1cm, minimum height=2cm, align=center
        },
        >=latex
      ]
      \node[block]                     (id)    {
        $\begin{aligned}
             \dot{\tilde{\zeta}} &= \mathcal{A}\tilde{\zeta}\\
             \tilde{v} &= \mathcal{C} \tilde{\zeta}
        \end{aligned}$
      };
      \node[block, right=0.8cm of id]    (opt)   {
        $\tau \dot{u} = \mathcal{H}_{w}(u,\tilde{v})$
      };
      \node[block, right=0.8cm of opt]   (plant) {
        $
        \begin{aligned}
            \dot x &= Ax + Bu + Ew \\
            y &= Cx
        \end{aligned}
        $
      };

      \node[above=0.4cm of opt] (w1) {$w$};
      \draw[->, thick] (w1) -- (opt.north);
  
      \node[above=0.4cm of plant] (w2) {$w$};
      \draw[->, thick] (w2) -- (plant.north);

      \draw[->, thick] (id.east)  -- node[black,midway,above] {$\tilde v$} (opt.west);
      \draw[->, thick] (opt.east) -- node[black,midway,above] {$u$}          (plant.west);
    \end{tikzpicture}

%% file: Figures/fig4.tex
\begin{tikzpicture}[
        block/.style = {
          rectangle, draw=black, fill=white,
          thick, 
          minimum width=0.4cm, minimum height=1cm, align=center,
          font=\scriptsize 
        },
        >=latex
      ]
      \node[block]                     (id)    {
        $\begin{aligned}
             \dot{\tilde{\zeta}} &= \mathcal{A} \tilde{\zeta}+ \mathcal{G} e_2 \\
             \tilde{v} &= C \tilde{\zeta}
        \end{aligned}$
      };
      \node[block, right=0.4cm of id]    (opt)   {
        $\tau \dot{u} = \mathcal{H}_w \bigl(u, \tilde v, e_2 \bigr)$
      };
      \node[block, right=0.4cm of opt]   (plant) {\small
        $
        \begin{aligned}
        \dot{x} &= Ax + Bu + Ew \\
            y &= Cx
        \end{aligned}
        $
      };
      \node[above=0.4cm of opt] (w1) {$w$};
      \draw[->, thick] (w1) -- (opt.north);
  
      \node[above=0.4cm of plant] (w2) {$w$};
      \draw[->, thick] (w2) -- (plant.north);

      \draw[->, thick] (id.east)  -- node[black,midway,above] {$\tilde v$} (opt.west);
      \draw[->, thick] (opt.east) -- node[black,midway,above] {$u$}          (plant.west);
    \draw[->,gray]
        (plant.south) -- ++(0,-0.5cm) node[gray,midway,right] {$e_2$} -| (opt.south);
    \draw[->,gray]
        (plant.south) -- ++(0,-0.5cm) -| (id.south);
    \end{tikzpicture}